\documentclass[%
 reprint,
preprint,
onecolumn,
 amsmath,amssymb,
 aps,
pra,
]{revtex4-2}
\usepackage{graphicx}
\usepackage{dcolumn}
\usepackage{bm}

\usepackage{subcaption} 

\usepackage[title]{appendix}
\usepackage{graphicx}
\usepackage{epstopdf}
\usepackage{amsmath,amssymb}
\usepackage{enumerate}
\usepackage{color}
\usepackage{cancel}
\usepackage{float}

\renewcommand{\thesection}{\arabic{section}}

\begin{document}

\title{Vortices in D-dimensional anisotropic Bose-Einstein condensates: dimensional perturbation theory with hypercylindrical symmetry}

\author{Maria Isabelle Fite}
 \affiliation{Department of Physics and Engineering Physics, University of Tulsa, Tulsa, Oklahoma 74104, USA.}
\author{B. A. McKinney}%
 \email{brett-mckinney@utulsa.edu}
\affiliation{%
 Tandy School of Computer Science, University of Tulsa, Tulsa, Oklahoma 74104, USA
}

\preprint{APS/123-QED}
\date{\today}
\begin{abstract} 
We investigate D-dimensional atomic Bose-Einstein condensates in a hypercylindrical trap with a vortex core along the z-axis and quantized circulation $\hbar m$. We analytically approximate the hypercylindrical Gross-Pitaevskii equation using dimensional perturbation theory with perturbation parameter $\delta=1/(D+2|m|-d)$, \textcolor{black}{where $d$ controls the contribution of kinetic energy at zeroth order}. We derive the zeroth-order ($\delta \to 0$) semiclassical approximations for the condensate energy, density, chemical potential, and critical vortex rotation speed in arbitrary dimensions. We investigate the effect of trap anisotropy on lower effective dimensionality and compute properties of vortices in higher dimensions motivated by the study of synthetic dimensions and holographic duality, where a higher-dimensional gravitational model corresponds to a lower-dimensional quantum model. In the zeroth-order approximation, we observe crossings between energy levels for different dimensions as a function of interaction strength and anisotropy parameters. 
\end{abstract}
\maketitle

\section{Introduction}



Experimental properties of magnetically trapped Bose-Einstein condensates (BECs) of dilute atomic gases at ultra-low temperatures are well described by the mean-field nonlinear Schrodinger equation known as the Gross-Pitaevskii equation (GPE) \cite{Fetter01}.  The current study focuses on BECs in arbitrary Cartesian dimensionality, which we model using the GPE in $D$-dimensional hypercylindrical coordinates.  We choose cylindrical coordinates because the most common anisotropic trap has axial symmetry. Using the trap anisotropy, BECs can have lower effective dimensionality, such as approximately 1D, 2D or isotropic 3D \cite{Gorlitz01}. We analyze the effect of anisotropy in the cylindrical geometry as a surrogate for effective lower dimension in an isotropic system.   

We also explore higher-dimensional vortex properties, which have potential applications in the emerging field of synthetic dimensions \cite{McCanna21, Boada12, Sugawa18}. In these experiments, internal degrees of freedom are manipulated so that they mathematically behave like extra external degrees of freedom such as extra spatial dimensions. The GPE has also been used with variable dimensionality to compare vortex cluster formation with predictions using a blackhole holographic model \cite{Yang24}. Holographic duality is a large-$D$ result from string theory in which anti-de Sitter spacetimes in $D$ spatial dimensions correspond to a conformal field theory on the boundary ($D-1$ spatial dimensions). The focus of the current study is general-$D$ approximations for the GPE quantum system, but the results could have applications in gravitational theory. 

Previously, we used dimensional perturbation theory (DPT) to study beyond mean field effects \cite{McKinney04-2}, and we used the large-$D$ limit to approximate solutions of the $D$-dimensional GPE \cite{McKinney02}. In both of these cases, we assumed an isotropic BEC and hyperspherical coordinates. In the current study, we extend the large-$D$ approximation of DPT for the isotropic GPE \cite{McKinney02} to the anisotropic GPE with $D$-dimensional hypercylindrical symmetry. We use $\delta=1/(D+2|m|-d)$ as our perturbation parameter, where $m$ is the vortex quantum number and $d$ can be thought of as the reference dimension for $D$. The parameter $d$ can also be used to control the kinetic energy contribution for $\delta \to 0$, which is a large-$D$ or large-$m$ approximation. In addition to keeping contributions from the kinetic energy, zeroth order also includes the full potential energy and nonlinear interaction from the GPE.  

We use the semiclassical approximation techniques in Ref. \cite{Sinha97} and extend their anisotropic trap vortex collective excitation results to $D$ dimensions. \textcolor{black}{We compare our analytical approximation with the variational approach for the anisotropic GPE for the ground state in Ref. \cite{Das02}.} We derive semiclassical approximations of key quantities, including the condensate density, energy, chemical potential, and critical vortex speed in D dimensions. We examine the effect of D on these quantities, and we observe crossings between energy levels for different dimensions as a function of interaction strength and anisotropy parameters. We also derive the $D$-dimensional Thomas-Fermi approximation. 

\section{The hypercylindrical GPE}\label{sec:GPE}
We assume a quantized vortex of a collection of ultracold trapped bosons, each with mass $M$, rotating about the $z$-axis in $D$-dimensions. This rotation results in hypercylindrical symmetry. 
\textcolor{black}{The hypercylindrical coordinate system is composed of one Cartesian coordinate, which we designate as $z$ and assume to be in the direction of the vortex core. The $z$ coordinate is orthogonal to a $D-1$ dimensional hyperspherical subspace ($D\ge2$) defined by a hyperradial coordinate $r_{\perp}$ (the radius of the $D-1$ dimensional hypersphere) and a set of $D-2$ angles. In the cases $D=3$ and $D=2$, the hypersphere reduces to a circle and line segment segment, respectively.  Essentially, we form the hypercylindrical Laplacian by adding $d^2/dz^2$ to a $(D-1)$-dimensional hyperspherical Laplacian. For hyperspherical and hypercylindrical coordinates, we use the convention in Ref. \cite{ GoodsonHerschbach1992}.} The time-independent GPE in $D\ge2$ hypercylindrical coordinates can then be written as
\begin{eqnarray}\label{eq:generalDGPE}
\biggl\{-\frac{\hbar^2}{2M} \biggl[\frac{1}{r_{\perp}^{D-2}} \frac{\partial}{\partial r_{\perp}} \biggl( r_{\perp}^{D-2} \frac{\partial}{\partial r_{\perp}} \biggr) + \frac{\partial^2}{\partial z^2} - \frac{L_{D-2}^2}{r_{\perp}^2} \biggr] +  \frac{1}{2}M\omega_{\perp}^2(r_{\perp}^2+\lambda^2z^2)&&\nonumber\\
+ \frac{u_D}{2}|\psi(\textbf{r})|^2 \biggr\} \psi(\textbf{r}) &&~= \mu \psi(\textbf{r}),
\end{eqnarray}
where $\omega_{\perp}$ is the transverse confinement oscillator frequency in the hyperspherical dimension perpendicular to $z$ and $\lambda = \frac{\omega_z}{\omega_{\perp}}$ is a measure of the anisotropy of the harmonic confinement. The hypercylindrical symmetry of the vortex fixes the $L_z$ component of the angular momentum to the vortex axis, and $L_{D-2}^2$ corresponds to the projection of the angular momentum. The eigenvalues of $L_{D-2}^2$ are $|m| (|m| + D - 3)$ \cite{Herschbach90}, 
where $|m| = 0, 1, 2, ...$. For the rotating BEC, $|m|$ is the magnitude of the projection of the angular momentum onto the $z$ axis and the sign of $m$ indicates the direction of rotation.  The wavefunction of the condensate, $\psi(\textbf{r})$, is normalized such that $\int|\psi(\textbf{r})|^2 = N$, where $N$ is the number of particles in the condensate.
The role of $L_{D-2}^2$ is analogous to the hypercylindrical symmetry of a $D$-dimensional $H_2^+$ molecular ion \cite{Herschbach86} or hydrogen in a magnetic field \cite{Baym96}.

The parameter $u_D$ is the coupling constant in $D$ dimensions, given by
\begin{eqnarray}\label{eq:generalDu}
u_D = \frac{\hbar^2}{M} \frac{2 \Omega_D}{\Gamma(D-2)} a^{D-2},
\end{eqnarray}
where $a$ is the s-wave scattering length and {$\Omega_D=2\pi^{\frac{D}{2}}/\Gamma(\frac{D}{2})$} is the angular integral over a $D$-dimensional sphere. Because of the gamma function in the denominator, the $D=2$ and $D=1$ limits for $u_D$ require careful treatment but do exist. These limits of the pseudopotential have been calculated in Ref. \cite{McKinney23} and Ref. \cite{Le19}, respectively.

Following the approach in Ref. \cite{McKinney02}, we introduce the Jacobian transformation of the wave function, $\phi=r^{(D-2)/2}\psi$, in order to eliminate the first derivatives from Eq. (\ref{eq:generalDGPE}):
\begin{eqnarray}\label{eq:transformedGPE}
\biggl\{-\frac{\hbar^2}{2M} \biggl[\frac{\partial^2}{\partial r_{\perp}^2} + \frac{\partial^2}{\partial z^2}
-  \frac{(D-2)(D-4)}{4r_{\perp}^2} -\frac{L_{D-2}^2}{r_{\perp}^2} \biggr] &&\nonumber\\ 
+ \frac{1}{2}M\omega_{\perp}^2(r_{\perp}^2+\lambda^2z^2) + \frac{u_D}{2}|\psi(\textbf{r})|^2 \biggr\} \phi(\textbf{r}) ~&&= \mu \phi(\textbf{r}).
\end{eqnarray}
Next, we introduce transverse oscillator units: $r_{\perp}=\ell_\perp \bar{r}_{\perp}$, $z=\ell_\perp \bar{z}$, $\mu = \hbar\omega_{\perp} \bar{\mu}$, $u_D=\hbar \omega_{\perp} \ell_\perp^D \bar{u}_D$ and $\psi(\textbf{r})=\ell_\perp^{-D/2} \bar{\psi}(\bar{\textbf{r}})$, where $\ell_\perp=\sqrt{\frac{\hbar}{M\omega_\perp}}$ is the oscillator length in the radial axis perpendicular to $z$. These transformations and substituting the eigenvalues of $L_{D-2}^2$ gives the following hypercylindrical GPE in transverse oscillator units:
\begin{eqnarray}\label{eq:oscillatorGPE}
\biggl\{ -\frac{1}{2} \biggl[\frac{\partial^2}{\partial \bar{r}_{\perp}^2} + \frac{\partial^2}{\partial \bar{z}^2} -  \frac{(D-2)(D-4)}{4\bar{r}_{\perp}^2} 
-\frac{|m|(|m|+D-3)}{\bar{r}_{\perp}^2} \biggr] &&\nonumber\\
+ \frac{1}{2}(\bar{r}_{\perp}^2+\lambda^2\bar{z}^2)
+ \frac{\bar{u}_D}{2}|\bar{\psi}(\bar{\textbf{r}})|^2 \biggr\} \phi(\textbf{r}) ~&&= \bar{\mu} \phi(\textbf{r}).
\end{eqnarray}


For the DPT approach, we define dimensionally scaled oscillator units: $\bar{r}_{\perp}=\kappa^{1/2}\hat{r}_{\perp}$, $\bar{z}=\kappa^{1/2}\hat{z}$, $\bar{\mu}=\kappa\hat{\mu}$, $\bar{u}_D=\kappa^{1-D}\hat{u}_D$, and $\bar{\psi}(\bar{\textbf{r}})=\kappa^{D/2}\hat{\psi}(\hat{\textbf{r}})$. The dimensionally scaled GPE becomes
\begin{eqnarray}\label{eq:scaledGPE}
\biggl\{ -\frac{1}{2}\delta^2 \biggl[\frac{\partial^2}{\partial \hat{r}_{\perp}^2} + \frac{\partial^2}{\partial \hat{z}^2} \biggr] + \frac{\textcolor{black}{1-2(3-d)\delta+(4-d)(2-d)\delta^2}}{8\hat{r}_{\perp}^2}
&&\nonumber\\ + \space \frac{1}{2}(\hat{r}_{\perp}^2 + \lambda^2\hat{z}^2)+\frac{\hat{u}_D}{2}|\hat{\psi}(\hat{\textbf{r}})|^2 &\biggr\} \phi(\textbf{r}) = \hat{\mu}\phi(\textbf{r}).
\end{eqnarray}
The perturbation parameter is defined as $\delta=1/\kappa$, where
\begin{eqnarray}\label{eq:kappa}
\kappa =D+2|m|-\textcolor{black}{d}.
\end{eqnarray}
The rationale behind the choice of $\kappa$ and $d$ is as follows. We assume the centrifugal numerator in Eq. (\ref{eq:oscillatorGPE}) equals a polynomial in $\kappa$: $(D-4)(D-2) + 4|m|(|m|+D-3) = \kappa^2 + b \kappa + c$. The constants $b$ and $c$ are independent of $D$ and $|m|$, while $\kappa$ is linear in $|m|$ and $D$. Using these constraints and plugging in $\kappa$ gives $b=-2(3\textcolor{black}{-d})$ and $c=(4\textcolor{black}{-d})(2\textcolor{black}{-d})$.
In Ref. \cite{McKinney02}, $\textcolor{black}{d}=2, 1, 0$ were used for different applications. \textcolor{black}{The parameter $d$ does not affect the mathematical form of the the zeroth-order ($\delta \to 0$) GPE density; however, it changes the weight of the centrifugal term through $\kappa$ in regular oscillator units. Thus, the parameter $d$ can be varied to increase or decrease the zeroth-order kinetic energy, where $d$ set to $d_{TF}=D+2|m|$ gives zero kinetic energy (Thomas Fermi). When $d=D$ for a given target dimension $D$, $\kappa = 2|m|$ and the only contribution to the zeroth-order kinetic energy is the centrifugal vortex term. This often makes the choice $d=D$ a natural `reference dimension' for many applications. For example, when comparing our zeroth order to the $D=3$ results in Ref. \cite{Sinha97}, where a centrifugal term is manually added to the Thomas-Fermi approximation, a natural choice is to let $d=3$. } 

\section{Zeroth order (large-$D$ or large-$|m|$) approximation}\label{sec:zerothOrderDensity}
We now take the zeroth-order $\delta \rightarrow 0$ limit of Eq. (\ref{eq:scaledGPE}), which can be thought of as a large-$|m|$ or large-$D$ limit. The zeroth-order approximation of the density is 
\begin{eqnarray}\label{eq:zerothOrderDensity}
|\bar{\psi}(\bar{\textbf{r}})|^2 \rightarrow \frac{1}{\bar{u}_D}\biggl(2\bar{\mu} - \bar{r}_\perp^2 - \lambda^2\bar{z}^2-\frac{\kappa^2}{4\bar{r}_\perp^2}\biggr),
\end{eqnarray}
where we reverted the dimensionally scaled oscillator (hat) units back to regular transverse oscillator (bar) units. In these units, the dependence on $D$ and $|m|$ is made explicit through $\kappa$. 
Note that for $D=3$ \textcolor{black}{and setting $d=3$} the centrifugal term becomes $|m|^2/\bar{r}_\perp^2$ in agreement with Ref. \cite{Sinha97}.
Since the chemical potential is part of the zeroth-order density, we use the general-$D$ normalization condition to find an expression for the chemical potential $\bar{\mu}$:
\begin{eqnarray}\label{eq:generalDNormalization}
\Omega_{D-1}\iint |\bar{\psi}(\bar{\textbf{r}})|^2 d\bar{z} \bar{r}_\perp^{D-2} d\bar{r}_\perp = N,
\end{eqnarray} 
where $\Omega_{D-1}$ is the angular integral corresponding to a $D-1$ dimensional sphere with $D-2$ angles. Using our zeroth-order approximation for $|\bar{\psi}(\bar{\textbf{r}})|^2$ (Eq.  \ref{eq:zerothOrderDensity}), the integral in Eq. (\ref{eq:generalDNormalization}) has an exact form in terms of the hypergeometric function ${}_2F_1$ (Appendix \ref{First_Appendix}). 

To find a simpler expression for $\bar{\mu}$ (compared to Appendix \ref{First_Appendix}), we define a perturbation parameter 
\begin{eqnarray}\label{eq:alpha}
    \textcolor{black}{\bar{\alpha}}=\kappa/(4\bar{\mu}),
\end{eqnarray}
which would be the same $\alpha$ used in Ref. \cite{Sinha97} for $D=3$ if we let $d=3$ in $\kappa$. We then define primed units $\bar{r}_\perp=\sqrt{2\bar{\mu}}r_\perp^\prime$, $\bar{z}=\sqrt{2\bar{\mu}}z^\prime$, and $|\psi^\prime(r^\prime, z^\prime)|^2=\frac{\Omega_{D-1}}{N} |\bar{\psi}(\bar{\textbf{r}})|^2$ to make the densities have unit norm for all $D$, which yields
\begin{eqnarray}\label{eq:zerothOrderDensityAlpha}
|\bar{\psi}(r^\prime, z^\prime)|^2 = \frac{\Omega_{D-1}}{N} \frac{2\bar{\mu}}{\bar{u}_D}\left(1 - r_\perp^{\prime2} - \lambda^2z^{\prime2} -\frac{\textcolor{black}{\bar{\alpha}}^2}{r^{\prime2}_\perp}\right).
\end{eqnarray}
The $z$ limits of integration in Eq. (\ref{eq:generalDNormalization}) can be found with the condition $|\bar{\psi}(\bar{\textbf{r}}^{\prime})|^2 \geq 0$:
\begin{eqnarray}
\label{eq:zNormUpperBound}
    z^{\prime}_{max} = \frac{\sqrt{1-r^{\prime2}_\perp-\frac{\textcolor{black}{\bar{\alpha}}^2}{r^{\prime2}_\perp}}}{\lambda}
\end{eqnarray}
\begin{eqnarray}
\label{eq:zNormLowerBound}
    z^{\prime}_{min} = -\frac{\sqrt{1-r^{\prime2}_\perp-\frac{\textcolor{black}{\bar{\alpha}}^2}{r^{\prime2}_\perp}}}{\lambda}.
\end{eqnarray}
With these substitutions and solving the $z^\prime$-integral exactly, the normalization condition yields
\begin{eqnarray}\label{eq:generalDNormalizationzInt}
\frac{4}{3\lambda} \frac{\Omega_{D-1} (2\bar{\mu})^\frac{D+2}{2}}{\bar{u}_D} \int (1 - r_\perp^{\prime2} -\frac{\textcolor{black}{\bar{\alpha}}^2}{r^{\prime2}_\perp})^{3/2} r_\perp^{\prime D-2} dr_\perp^\prime = N.
\end{eqnarray}
The $r^\prime_\perp$ limits of integration at the edge of the condensate are then
\begin{eqnarray}
\label{eq:normLowerBound}
    r^{\prime}_{\perp min} = \sqrt{\frac{1-\sqrt{1-4\textcolor{black}{\bar{\alpha}}^2}}{2}}.
\end{eqnarray}
\begin{eqnarray}
\label{eq:normUpperBound}
    r^{\prime}_{\perp max} = \sqrt{\frac{1+\sqrt{1-4\textcolor{black}{\bar{\alpha}}^2}}{2}}.
\end{eqnarray}
The value $r^{\prime}_{\perp min}$ is the inner radius of the condensate and can be thought of as the radius of the vortex core, while $r^{\prime}_{\perp max}$ is the outer radius of the condensate. Note that $r^{\prime}_{\perp min}$ and $r^{\prime}_{\perp max}$ are only real for $\textcolor{black}{\bar{\alpha}}^2 \le \frac{1}{4}$, which yields a lower bound for the chemical potential, $\bar{\mu} \ge \kappa/2$.  

\begin{figure}[H]
\includegraphics{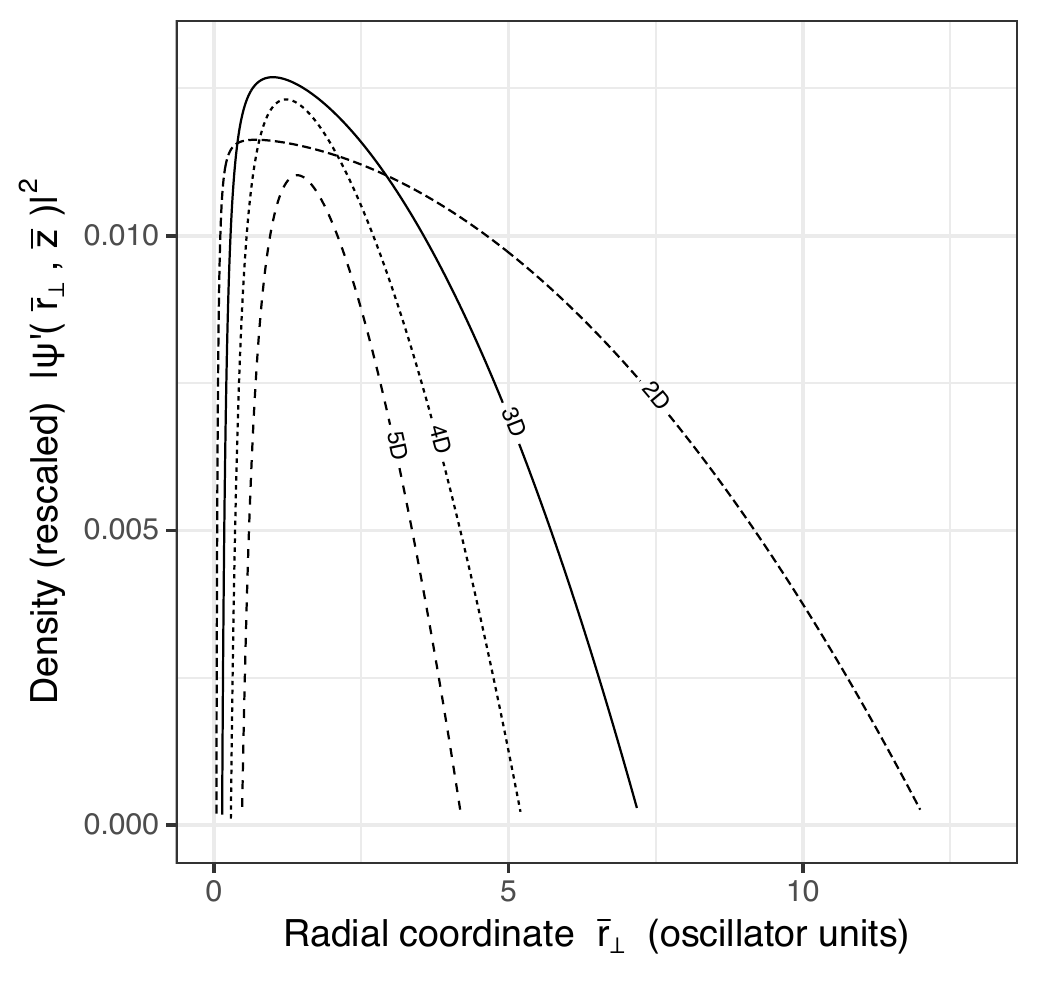}
\caption{Side view cross section of zeroth-order density $|\psi^\prime(\bar{\textbf{r}})|^2$ of Bose-Einstein condensates with different numbers of dimensions using Eqs.(\ref{eq:zerothOrderDensity}) and (\ref{eq:generalDMu}). The density $|{\psi^\prime}(\bar{\textbf{r}})|^2 =\Omega_{D-1} |\bar{\psi}(\bar{\textbf{r}})|^2/N$ is scaled so that $\iint |{\psi^\prime}(\bar{\textbf{r}})|^2 d\bar{z} \bar{r}_\perp^{D-2} d\bar{r}_\perp = 1$ for each $D$. We use $N=1000$ atoms, vortex number $m=1$, anisotropy $\lambda=4/3$ and interaction strength $\bar{u}_D \approx 25.1$, which corresponds to scattering length $\bar a=1$ in oscillator units. We use $\kappa=D+2|m|-3$. The density becomes more squeezed with increasing $D$.}  \label{fig:densityAtDiffD}
\end{figure}

\begin{figure}[H]
\includegraphics{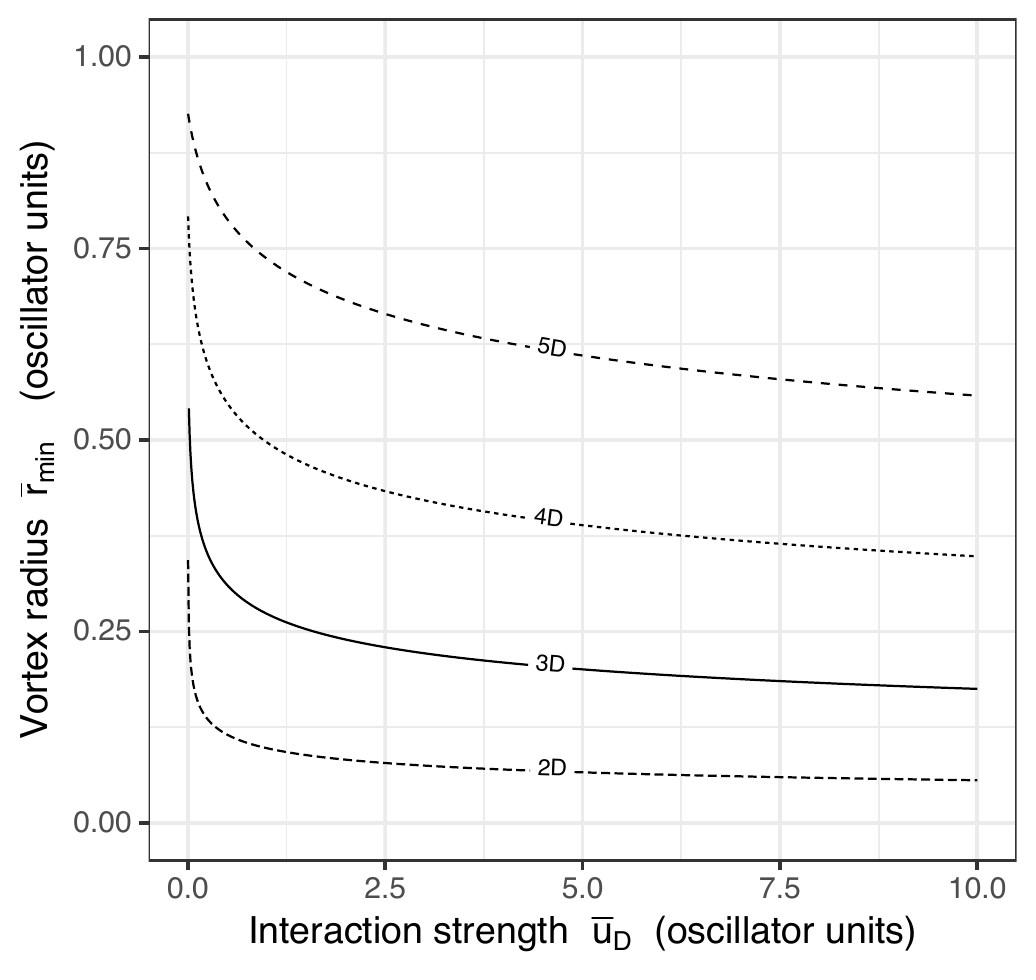}
\caption{Inner (vortex core) radius $r^{\prime}_{\perp min}$ (from Eq. \ref{eq:normLowerBound} and $\bar{\mu}$ from Eq. \ref{eq:generalDMu}) as a function of the interaction strength, $\bar{u}_D$. We use $N=1000$ atoms, vortex number $m=1$, and isotropic condensates with $\lambda=1$ We use $\kappa=D+2|m|-3$.} \label{fig:rminVersusu}
\end{figure}

Finally, expanding the normalization integrand in Eq. (\ref{eq:generalDNormalizationzInt}) to second order in $\textcolor{black}{\bar{\alpha}}$, we obtain an expression that can be solved numerically for the chemical potential $\bar{\mu}$:
\begin{eqnarray}
\label{eq:generalDMu}
\frac{4\Omega_{D-1}(2\bar{\mu})^\frac{D+2}{2}}{3\lambda\bar{u}_D} \biggl[ \frac{\textcolor{black}{\bar{\alpha}}^{D-1}(D+3)}{2(D-1)(D-3)}
-\frac{3\textcolor{black}{\bar{\alpha}}^2\sqrt{\pi}\Gamma(\frac{D-1}{2})}{4(D-3)\Gamma(\frac{D}{2})} + \frac{\sqrt{\pi}\Gamma(\frac{D+1}{2})}{2(D-1)\Gamma(\frac{D+2}{2})} &&\nonumber\\
-\frac{\sqrt{\pi}\Gamma(\frac{D+3}{2})}{2(D+1)\Gamma(\frac{D+4}{2})} \biggr] &&~\approx N.
\end{eqnarray}
Determining the density (Eq. \ref{eq:zerothOrderDensityAlpha}) requires solving Eq. (\ref{eq:generalDMu}) for $\bar{\mu}$ and restricting $r'_{\perp}$ to its lower and upper limits (Eqs. \ref{eq:normLowerBound} and \ref{eq:normUpperBound}). As $D$ increases, the vortex core is pushed outward due to the centrifugal term in the large-$D$ density, and at the same time, the outer surface of the condensate is pulled inward along the $r_\perp$ dimension (Fig. \ref{fig:densityAtDiffD}). As the interaction strength $\bar{u}_D$ increases, the radius of the vortex core $r^{\prime}_{\perp min}$ decreases rapidly at first and then flattens (Fig. \ref{fig:rminVersusu}). 

Despite the appearance of $\frac{1}{D-3}$ terms, Eq. (\ref{eq:generalDMu}) has a $D\rightarrow3$ limit: 
\begin{eqnarray}\label{eq:3DMu} \frac{8\pi(2\bar{\mu})^\frac{5}{2}}{3\lambda\bar{u}_3} \biggl[ \frac{1}{5}+\frac{3\textcolor{black}{\bar{\alpha}}^2}{2} \biggl( \ln{\frac{\textcolor{black}{\bar{\alpha}}}{2}} + \frac{2}{3} \biggr) \biggr] \approx N,
\end{eqnarray}
in agreement with the form in Ref. \cite{Sinha97} and exact agreement when $d=3$. Similarly, \textcolor{black}{for $D=2$, we find}
\textcolor{black}{
\begin{eqnarray}\label{eq:2DMu} 
\frac{8(2\bar{\mu})^2}{3 \lambda \bar{u}_2}\bigg[ \frac{3\pi\textcolor{black}{\bar{\alpha}}^2}{4}- \frac{5 \textcolor{black}{\bar{\alpha}}}{2} + \frac{3\pi}{16}\bigg]  \approx N,
\end{eqnarray}}
\textcolor{black}{which can be solved for $\bar{\mu}$:} 
\textcolor{black}{
\begin{eqnarray}\label{eq:2DMuSolved}
\bar{\mu} = \frac{10|\kappa|}{6 \pi}\pm\sqrt{\left(\frac{25}{9 \pi ^2}-\frac{1}{4}\right) \kappa ^2+\left(\frac{N\lambda  \bar{u}_2}{2\pi}\right)}.
\end{eqnarray}}
\textcolor{black}{The $D\to1$ limit exists, but this is technically not a physical limit in the hypercylidrical coordinates. }

Higher-order expansions in $\textcolor{black}{\bar{\alpha}}$ (beyond Eq. \ref{eq:generalDMu}) accumulate additional poles at odd integer dimensions. The second-order expansion has no poles, while the fourth-order has a pole at $D=3$ and the sixth-order has poles at $D=3, 5$ (see Appendix \ref{Second_Appendix}). For our purposes, the second-order expansion in $\textcolor{black}{\bar{\alpha}}$ is a reasonable approximation, while having the advantage of a finite solution for $D=3$. In general, we hypothesize that the $n$th-order expansion in $\textcolor{black}{\bar{\alpha}}$ has poles at $D=3, 5, 7,...,2n-1$. So, the highest expansion possible for a given odd $D$, while avoiding a pole at $D$, is the $(D-1)$-order expansion. There are no poles at even $D$ regardless of the order of expansion.

We can use Eq. (\ref{eq:generalDMu}) to derive the general-$D$ Thomas-Fermi approximation for the ground state, where the kinetic energy term is neglected completely. 
In general, the kinetic energy contribution to the zeroth-order density (Eq. 
 \ref{eq:zerothOrderDensity}) will be nonzero even when $|m|=0$ because the number of dimensions $D$ and the parameter $d$ also contribute to $\kappa$ in the centrifugal term. \textcolor{black}{However, when the parameter $d$ is chosen to be $d_{TF} = 2|m| + D$, the remaining kinetic energy from the vortex in Eq. 
 (\ref{eq:zerothOrderDensity}) is forced to zero, and the zeroth-order density reduces to Thomas-Fermi}. The general-$D$ Thomas-Fermi chemical potential follows from Eq. (\ref{eq:generalDMu}) by letting $d=d_{TF}$, which results in $\kappa$ and $\alpha$ becoming zero:
\begin{eqnarray}\label{eq:generalDMuThomasFermi}
2\bar{\mu}_{TF}=\left(\frac{(D-1)}{\Omega_{D-1}}\frac{N \lambda \bar{u}_D \Gamma(\frac{D+4}{2})}{\sqrt{\pi} \Gamma (\frac{D+1}{2})}\right)^{\frac{2}{D+2}}.
\end{eqnarray}
This gives the known $D=3$ Thomas Fermi result (e.g., Ref. \cite{Sinha97}) and gives the correct results for $D=2$. \textcolor{black}{Despite not being a physical limit in our formalism, the $D\to1$ limit $(D-1)/\Omega_{D-1} \to 1$ in Eq. (\ref{eq:generalDMuThomasFermi}) leads to the correct $1D$ Thomas Fermi result. One could also derive $\bar{\mu}_{TF}$ from Eq. (\ref{eq:zerothOrderDensity}) and using the general-$D$ normalization condition (Eq. \ref{eq:generalDNormalization})}. Our general-$D$ and $m$ implicit formula for the chemical potential (Eq. \ref{eq:generalDMu}) can then be expressed in terms of the $D$-dimensional Thomas-Fermi chemical potential $\bar{\mu}_{TF}$:
\begin{eqnarray}
\label{eq:combinationMu}
\frac{4\Omega_{D-1}(2\bar{\mu})^\frac{D+2}{2}}{3\lambda\bar{u}_D} \biggl[ \frac{\textcolor{black}{\bar{\alpha}}^{D-1}(D+3)}{2(D-1)(D-3)} -\frac{3\textcolor{black}{\bar{\alpha}}^2\sqrt{\pi}\Gamma(\frac{D-1}{2})}{4(D-3)\Gamma(\frac{D}{2})} - \frac{\sqrt{\pi}\Gamma(\frac{D+3}{2})}{2(D+1)\Gamma(\frac{D+4}{2})}  \biggr] &&\nonumber\\
+ \frac{N(D+2)}{3}\biggr(\frac{\bar{\mu}}{\bar{\mu}_{TF}}\biggl)^{\frac{D+2}{2}} &&~\approx N,
\end{eqnarray}
where recall $\textcolor{black}{\bar{\alpha}} = \kappa/(4\bar{\mu}) = (D + 2|m| - d)/(4\bar{\mu})$ contains the vortex quantum number information. \textcolor{black}{Again, when $d=d_{TF}$, $\alpha$ becomes zero and one recovers $\mu_{TF}$ from Eq. (\ref{eq:combinationMu})}.

\begin{figure}[H]
\includegraphics{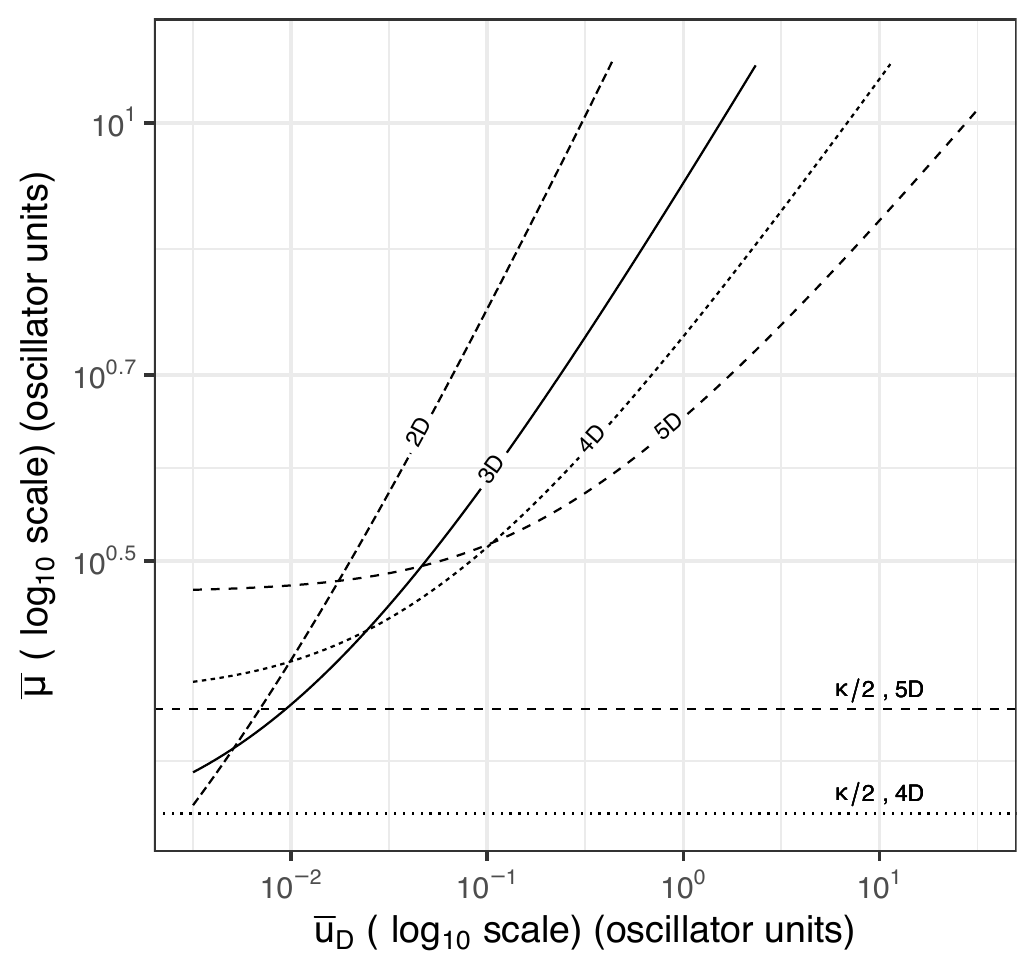}
\caption{Chemical potential $\bar{\mu}$ of different $D$ dimensional condensates as a function of the interaction strength, $\bar{u}_D$, plotted on a log-log scale from Eq. (\ref{eq:generalDMu}). We let $N=1000$, $m=1$, and $\lambda=1$. The horizontal lines are the lower bounds of the chemical potential ($\bar\mu=\kappa/2$) for $D=4$ and $D=5$. For weak interaction (left), $\bar{\mu}$ increases with increasing $D$, similar to a harmonic oscillator. For strong interaction (right), $\bar{\mu}$ decreases with increasing $D$ due to the Thomas-Fermi $D$ dependence. This leads to dimensional crossings. We use $\kappa=D+2|m|-3$. } \label{fig:muVersusu}
\end{figure}

Next, we consider the effect on the zeroth-order chemical potential of sweeping the parameters $\bar{u}_D$ (interaction strength) and $\lambda$ (anisotropy) for $D=2,3,4,5$. We refer to the curve of chemical potentials for a given $D$ as the $\bar{\mu}_D$-curve. \textcolor{black}{Each $\bar{\mu}_D$-curve increases with anisotropy or interaction strength for a given D. For fixed weak interaction strength, the $\bar{\mu}_D$-curves increase with increasing $D$ (i.e., lower dimensions have lower chemical potential, see far left side of Fig. \ref{fig:muVersusu}), which is expected based on the known $D$-dependence for the energy of a (non-interacting) harmonic oscillator. However, as the interaction strength $\bar{u}_D$ increases, the $\bar{\mu}_D$-curves cross each other, and for large interaction, the $\bar{\mu}_D$-curves are ordered by decreasing $D$ (i.e., higher dimensions have lower chemical potential, see right side of Fig. \ref{fig:muVersusu}), which is due to the dominance of the general Thomas-Fermi approximation (Eq. \ref{eq:generalDMuThomasFermi}) and its $D$-dependence of $2/(D+2)$ in the exponent. For weak interaction, the Thomas-Fermi contribution goes to zero and the harmonic oscillator energy dominates.} 

\begin{figure}[H]
\includegraphics{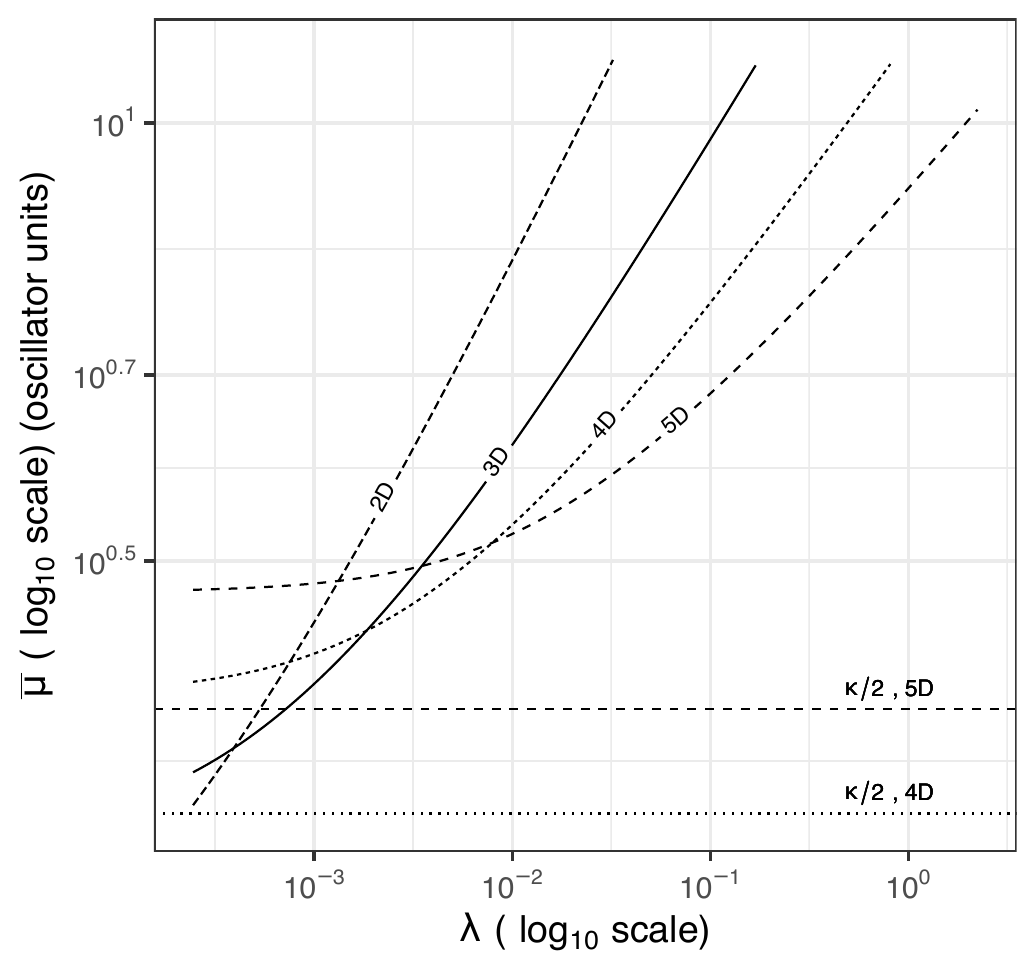}
\caption{Chemical potential $\bar{\mu}$ of different $D$ dimensional condensates as a function of the trap anisotropy, $\lambda$, plotted on a log-log scale from Eq. (\ref{eq:generalDMu}). We let $N=1000$, $m=1$, and $\bar{u}_D=25.1$. The horizontal lines are the lower bounds of the chemical potential ($\bar\mu=\kappa/2$) for $D=4$ and $D=5$. We use $\kappa=D+2|m|-3$. For large $lambda$ the Thomas-Fermi limit has a stronger contribution and we observe dimensional crossings similar to Fig. \ref{fig:muVersusu}. } \label{fig:muVersusLambda}
\end{figure}

The $\bar{\mu}_D$-curves show the same pattern as interaction strength when $\lambda$ anisotropy parameter is swept (Fig. \ref{fig:muVersusLambda}). This similarity is because the parameters $\bar{u}_D$ and $\lambda$ play the same role in Eq. (\ref{eq:generalDMu}). The plots also include the $\bar{\mu} = \kappa/2$ (horizontal lines in Figs. \ref{fig:muVersusu} and \ref{fig:muVersusLambda}), which come from the condition that the hyper-radii (Eqs. \ref{eq:normUpperBound} and \ref{eq:normLowerBound}) be real ($\bar{\mu} \ge \kappa/2$). This is a lower bound on $\bar{\mu}$ that depends on the number of dimensions $D$ and the magnitude of the angular momentum quantum number $m$. As $D$ or $|m|$ increases, the lower bound on $\bar{\mu}$ will also increase. The computed values of $\bar{\mu}$ stay above this lower bound even for small interactions.  

\section{Energy and critical velocity in D Dimensions: Zeroth order}\label{sec:energy}
The general-D energy functional of the condensate is given by
\begin{eqnarray} \label{eq:energyDef}
E = \int d^Dr\biggl(
\frac{\hbar^2}{2M}|\nabla \psi(\textbf{r})|^2 + V_{trap}(\textbf{r})|\psi(\textbf{r})|^2 + \frac{u_D}{4}|\psi(\textbf{r})|^4\biggr),
\end{eqnarray}
where $N \gg 1$, and which can be written in terms of the chemical potential as
\begin{eqnarray}\label{eq:generalDEnergyInt}
\frac{\bar{E}}{N} = \bar{\mu} - \frac{\bar{u}_D}{4N} \int d^D\bar{r} |\bar{\psi}(\bar{\textbf{r}})|^4.
\end{eqnarray}
Using our zeroth-order approximation for $\psi$ in Eq. (\ref{eq:zerothOrderDensity}), and expanding the integrand of Eq. (\ref{eq:generalDEnergyInt}) to 2nd order in $\textcolor{black}{\bar{\alpha}}$, we obtain the general-$D$ energy per particle:
\begin{eqnarray}\label{eq:generalDEnergy}
\frac{\bar{E}}{N} \approx~&& \bar{\mu} - \frac{(2\bar{\mu})^{\frac{D+4}{2}}\Omega_{D-1}}{30\lambda N \bar{u}_D} \biggl[ \frac{4\textcolor{black}{\bar{\alpha}}^{D-1}(3D+1)}{(D-1)(D-3)}
- \frac{15\textcolor{black}{\bar{\alpha}}^2\sqrt{\pi}\Gamma(\frac{D-1}{2})}{(D-3)\Gamma(\frac{D+2}{2})} + \frac{6\sqrt{\pi}\Gamma(\frac{D+1}{2})}{(D-1)\Gamma(\frac{D+4}{2})} \nonumber\\&&- \frac{6\sqrt{\pi}\Gamma(\frac{D+3}{2})}{(D+1)\Gamma(\frac{D+6}{2})} \biggr].
\end{eqnarray}
Note that $\bar{\mu}$ and $\textcolor{black}{\bar{\alpha}}$ depend on the vortex quantum number $m$. In the $D\rightarrow3$ limit, Eq. (\ref{eq:generalDEnergy}) reduces to 
\begin{eqnarray}\label{eq:3DEnergy}
\frac{\bar{E}}{N} \approx \bar{\mu} - \frac{8\pi(2\bar{\mu})^{7/2}}{15\lambda N \bar{u}_3} \biggl[\frac{1}{7} + \frac{5\textcolor{black}{\bar{\alpha}}^2}{2} \biggl(\ln{\frac{\textcolor{black}{\bar{\alpha}}}{2}} + \frac{17}{15} \biggr) \biggr],
\end{eqnarray}
in agreement with Ref. \cite{Sinha97}. 

The energy expression in Eq. (\ref{eq:generalDEnergy}) allows us to approximate the critical angular velocity $\bar{\Omega}_c$ to produce a quantized vortex with quantum number $m$ of a BEC in $D$ dimensions: 
\begin{eqnarray}\label{eq:critVelocity}
\bar{\Omega}_c(|m|) = \frac{1}{|m|} \biggl[ \bar{E}(|m|)/N - \bar{E}(0)/N \biggl],
\end{eqnarray}
where  $\bar{\Omega}_c$ and $\bar{E}/N$ are in transverse oscillator units. 
Substituting our energy approximation (Eq. \ref{eq:generalDEnergy}) into the critical velocity equation (Eq. \ref{eq:critVelocity}) we find
\begin{eqnarray}
\bar{\Omega}_c(|m|) \approx&&~ \frac{1}{|m|}\frac{\Omega_{D-1}}{30 \lambda \bar{u}_DN}\Biggl[(2\bar{\mu}(0))^{\frac{D+4}{2}}  \Biggl(\frac{4 (3 D+1)\left(\frac{\kappa(0)}{4\bar{\mu}(0)}\right){}^{D-1}}{(D-3) (D-1)}\nonumber\\&&
-\frac{15 \sqrt{\pi} (\frac{\kappa(0)}{4\bar{\mu}(0)})^2 \Gamma \left(\frac{D-1}{2}\right)}{(D-3) \Gamma \left(\frac{D+2}{2}\right)}+ \frac{6 \sqrt{\pi }\Gamma \left(\frac{D+1}{2}\right)}{(D-1) \Gamma \left(\frac{D+4}{2}\right)}-\frac{6 \sqrt{\pi }\Gamma \left(\frac{D+3}{2}\right)}{(D+1) \Gamma \left(\frac{D+6}{2}\right)}\Biggl)\nonumber\\&&
-(2 \bar{\mu}(|m|))^{\frac{D+4}{2}} \Biggl(\frac{4(3 D+1) \left(\frac{\kappa}{4\bar{\mu}(|m|)}\right){}^{D-1}}{(D-3) (D-1)}-\frac{15 \sqrt{\pi} (\frac{\kappa}{4\bar{\mu}(|m|)})^2 \Gamma \left(\frac{D-1}{2}\right)}{(D-3) \Gamma \left(\frac{D+2}{2}\right)}\nonumber\\&& 
+ \frac{6 \sqrt{\pi } \Gamma \left(\frac{D+1}{2}\right)}{(D-1) \Gamma \left(\frac{D+4}{2}\right)}-\frac{6 \sqrt{\pi } \Gamma \left(\frac{D+3}{2}\right)}{(D+1) \Gamma \left(\frac{D+6}{2}\right)}\Biggr)\Biggr] + \frac{\bar{\mu}(|m|)-\bar{\mu}(0)}{|m|},
\end{eqnarray}
where $\Omega_c(|m|)$ is the critical velocity for the vortex number $m$, $\bar{\mu}(0)$ is the ground-state chemical potential when $m=0$, $\bar{\mu}(|m|)$ is the excited-state chemical potential for the $L_z$ quantum number, $
\kappa = D + 2|m| - d$ and $\kappa(0) = D - d$. As in Eq. (\ref{eq:generalDEnergy}), $\lambda=\frac{\omega_z}{\omega_{\perp}}$ is the anisotropy parameter, $\bar{u}_D$ is the interaction strength and $N$ is the number of particles. For this approximation, we used zeroth order (in $\delta=1/\kappa$) and then expanded in $\textcolor{black}{\bar{\alpha}}$. The critical velocity is affected by the number of dimensions in the same way as the expression for energy.

In the limit as $D \rightarrow 3$, the critical velocity becomes
\begin{eqnarray}
\bar{\Omega}_c(|m|) \approx&&~ \frac{1}{|m|}\frac{8\pi}{15\lambda N \bar{u}_3}\Biggl[(2\bar{\mu}(0))^{7/2} \biggl(\frac{1}{7} + \frac{5}{2} \left(\frac{\kappa(0)}{4\bar{\mu}(0)}\right)^2 \biggl(\ln{\frac{\kappa(0)}{8\bar{\mu}(0)}} + \frac{17}{15} \biggr) \biggr) \nonumber\\&&
- (2\bar{\mu}(|m|))^{7/2} \biggl(\frac{1}{7} + \frac{5}{2} \left(\frac{\kappa}{4\bar{\mu}(|m|)}\right)^2 \biggl(\ln{\frac{\kappa}{8\bar{\mu}(|m|)}} + \frac{17}{15} \biggr) \biggr)\Biggr] + \frac{\bar{\mu}(|m|)-\bar{\mu}(0)}{|m|}.
\end{eqnarray}
Letting $d=3$ for the reference dimension, $\bar{\alpha} = (D+2|m|-3)/(4\bar{\mu}(|m|))$ becomes $|m|/(2\bar{\mu}(|m|))$ as $D \to 3$.
Taking the limit as $\bar{\alpha}$ goes to 0 ($\bar{\mu} \gg |m|$), the $3D$ critical velocity becomes
\begin{eqnarray}\label{eq:critVelocity3DSmallAlpha}
\bar{\Omega}_c \approx \frac{|m|}{12 \bar{\mu}(0)}\biggl[15 \ln\frac{4\bar{\mu}(0)}{|m|}-17\biggr],
\end{eqnarray}
in agreement with the $D=3$ critical velocity approximation in Ref. \cite{Sinha97}.


\section{Bose-Einstein Condensates in Four Dimensions: Synthetic Dimensions}\label{sec:4DCase}

BECs with $D \geq 3$ can be experimentally studied using ``synthetic dimensions,'' where internal degrees of freedom are used to simulate additional dimensions. 
Here we consider a $4D$ BEC with a vortex in hypercylindrical symmetry. 
Taking the limit as $D \rightarrow 4$ in Eqs. (\ref{eq:zerothOrderDensity}, \ref{eq:generalDMu}, \ref{eq:generalDEnergy}, and \ref{eq:critVelocity}) we can estimate the density, energy, chemical potential and critical velocity of the $4D$ BEC. The density using our zeroth-order approximation becomes 
\begin{eqnarray}\label{eq:zerothOrderDensity4D}
|\bar{\psi}(\bar{\textbf{r}})|^2 = \frac{1}{\bar{u}_4}\left(2\bar{\mu} - \bar{r}_\perp^2 - \lambda^2\bar{z}^2-\frac{\kappa^2}{4\bar{r}_\perp^2}\right),
\end{eqnarray}
where $\bar{u}_4 = 4 \pi^2 a^2$ is the coupling constant in $4$ dimensions. The addition of one dimension increases the power of the scattering length in $\bar{u}_d$ ($\bar{a}^2$ instead of $\bar{a}$) and  mimics the effect of increasing $|m|$ by 1/2, thus increasing the centrifugal kinetic energy contribution. 
In the $4D$ case, the normalization condition leads to
\begin{eqnarray}\label{eq:4DMu}
\frac{16\pi(2\bar{\mu})^3}{3\lambda\bar{u}_4} \biggl[ \frac{7\textcolor{black}{\bar{\alpha}}^{3}}{6} -\frac{3\pi\textcolor{black}{\bar{\alpha}}^2}{8} +
\frac{\pi}{32} \biggr]\approx N,
\end{eqnarray}
where $\textcolor{black}{\bar{\alpha}} = \kappa/(4\bar{\mu}) = (4+2|m|-d)/(4\bar{\mu})$. The energy per particle is 
\begin{eqnarray}\label{eq:4DEnergy}
\frac{\bar{E}}{N} \approx \bar{\mu} - \frac{2\pi(2\bar{\mu})^{4}}{15\lambda N \bar{u}_4} \biggl[ \frac{52\textcolor{black}{\bar{\alpha}}^{3}}{3} - \frac{15\pi\textcolor{black}{\bar{\alpha}}^2}{4} + \frac{5\pi}{32} \biggr],
\end{eqnarray}
and the critical velocity is
\begin{eqnarray}\label{eq:4DcritVelocity}
\bar{\Omega}_c(|m|) \approx&&~ \frac{1}{20|m|}\Biggl[-15\bar{\mu}(0) +(69\bar{\mu}(0)-45\pi\bar{\mu}^2(0))x^{-1} + 2^{-2/3}15x^{1/3} \nonumber\\&&
-104\kappa^3x^{-2/3}+ 2^{-2/3}30\kappa^2\left(\frac{3\pi}{2}\right)^{2/3}x^{-1/3}\Biggr].
\end{eqnarray}
where the variable $x = 7 - 9\pi \bar{\mu}(0) + 12\pi \bar{\mu}^3(0) = 9\bar{u}_4\lambda N/\pi$ and  $\bar{\mu}(0)$ is the $4D$ chemical potential for $m=0$.

\section{Anisotropy and lower effective isotropic dimensions}\label{sec:nonIntegerCases}

\begin{figure}
\includegraphics{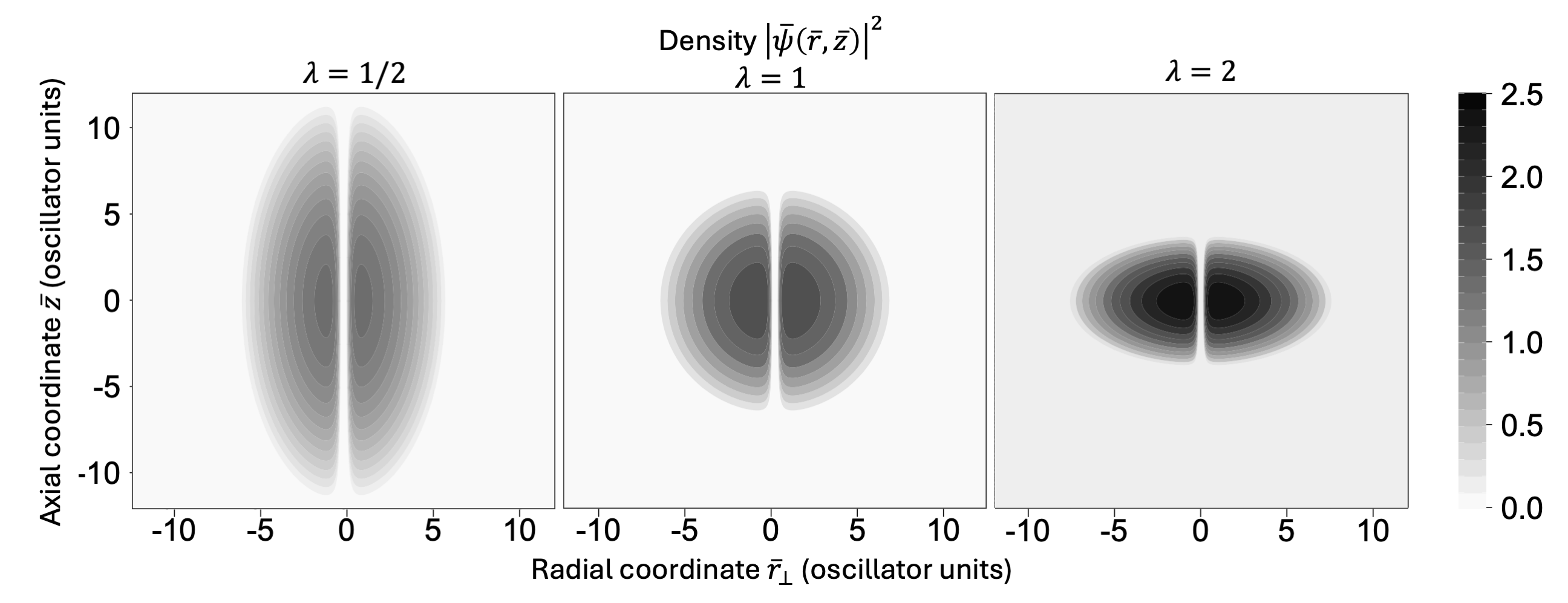}
\caption{Contour plots of the zeroth-order density $|\bar{\psi}(\bar{\textbf{r}})|^2$ (Eq.\ref{eq:zerothOrderDensity}), for $D=3$ condensates with three values of the anisotropy parameter, $\lambda=\frac{\omega_z}{\omega_{\perp}}$. Each density is normalized such that $\iint |{\bar{\psi}}(\bar{\textbf{r}})|^2 d\bar{z} \bar{r}_{\perp}^{D-2} d\bar{r}_{\perp} = N$. The coordinates $\bar{r}_\perp$ and $\bar{z}$ are in scaled transverse oscillator units. The condensates have an $m=1$ vortex along the $z$-axis, $N=1000$ atoms, and interaction strength $\bar{u}_D \approx 25.1$.}\label{fig:lambdaPanel}
\end{figure}

Recall that the anisotropy of a hypercylindrical condensate is controlled by $\lambda = \frac{\omega_z}{\omega_{\perp}}$, the ratio of confinement oscillator frequencies. Our zeroth order density (Eq. \ref{eq:zerothOrderDensity}) captures condensate features due to $\lambda$, such as stretching along the $z$-axis for $\lambda \ll 1$, making the BEC more 1$D$ (cigar geometry), while $\lambda \gg 1$ compresses along the $z$-axis, making the BEC more of a $2D$ disk (Fig. \ref{fig:lambdaPanel}) or a $D-1$ hyper-disk embedded in $D$ dimensions. \textcolor{black}{Here we further explore the role of $\lambda$ on lower effective dimension embedded in $D=3$, and we compare numerically with the two accurate variational expressions (small- and large-$\lambda$ limits) for $m=0$ and $D=3$ from Ref. \cite{Das02}.} 


\begin{figure}
\centering
\begin{subfigure}{.5\textwidth}
  \centering
  \includegraphics[width=1.0\linewidth]{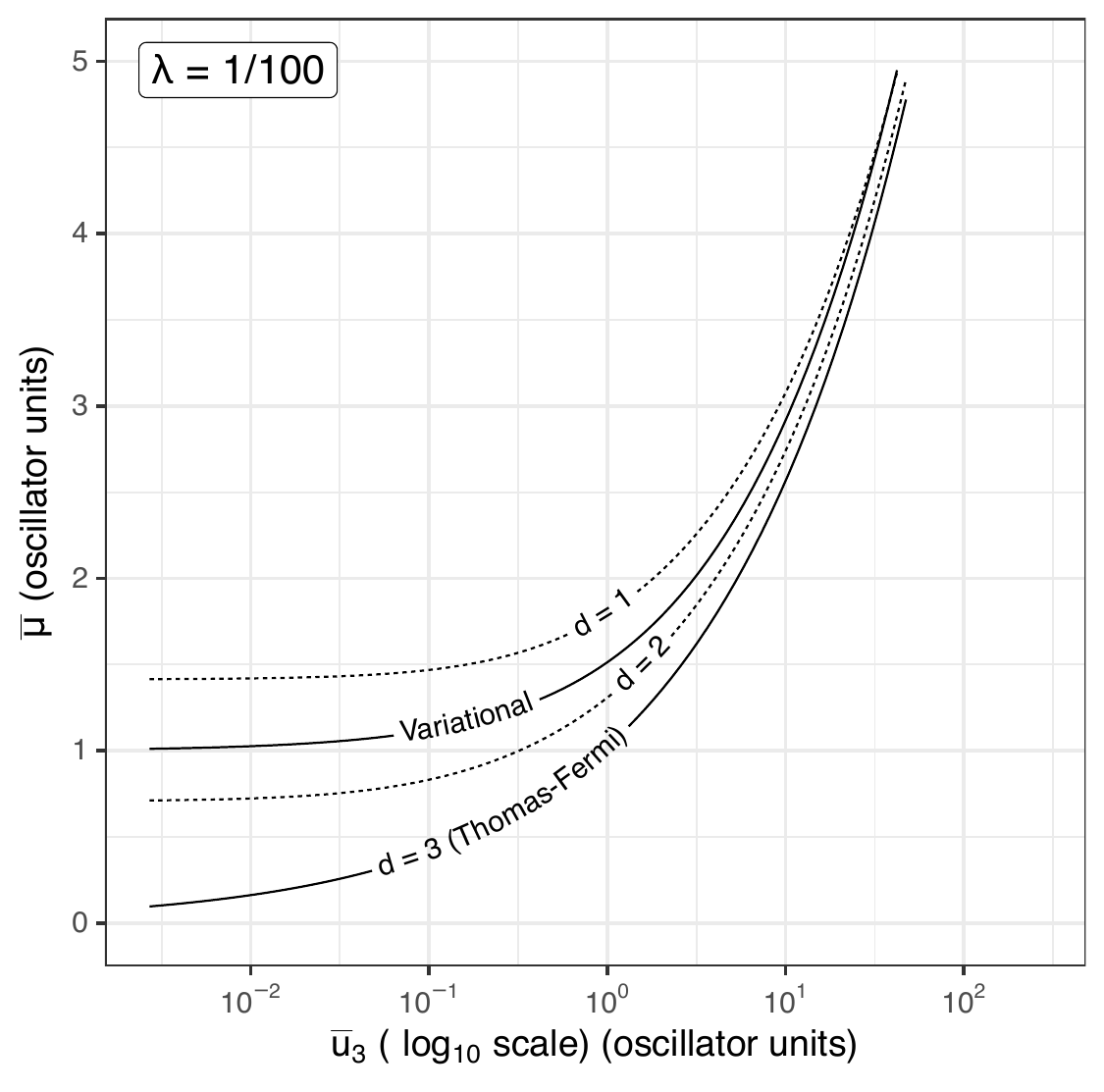}
  \label{fig:cigarGeometryMuComparison}
\end{subfigure}%
\begin{subfigure}{.5\textwidth}
  \centering
  \includegraphics[width=1.0\linewidth]{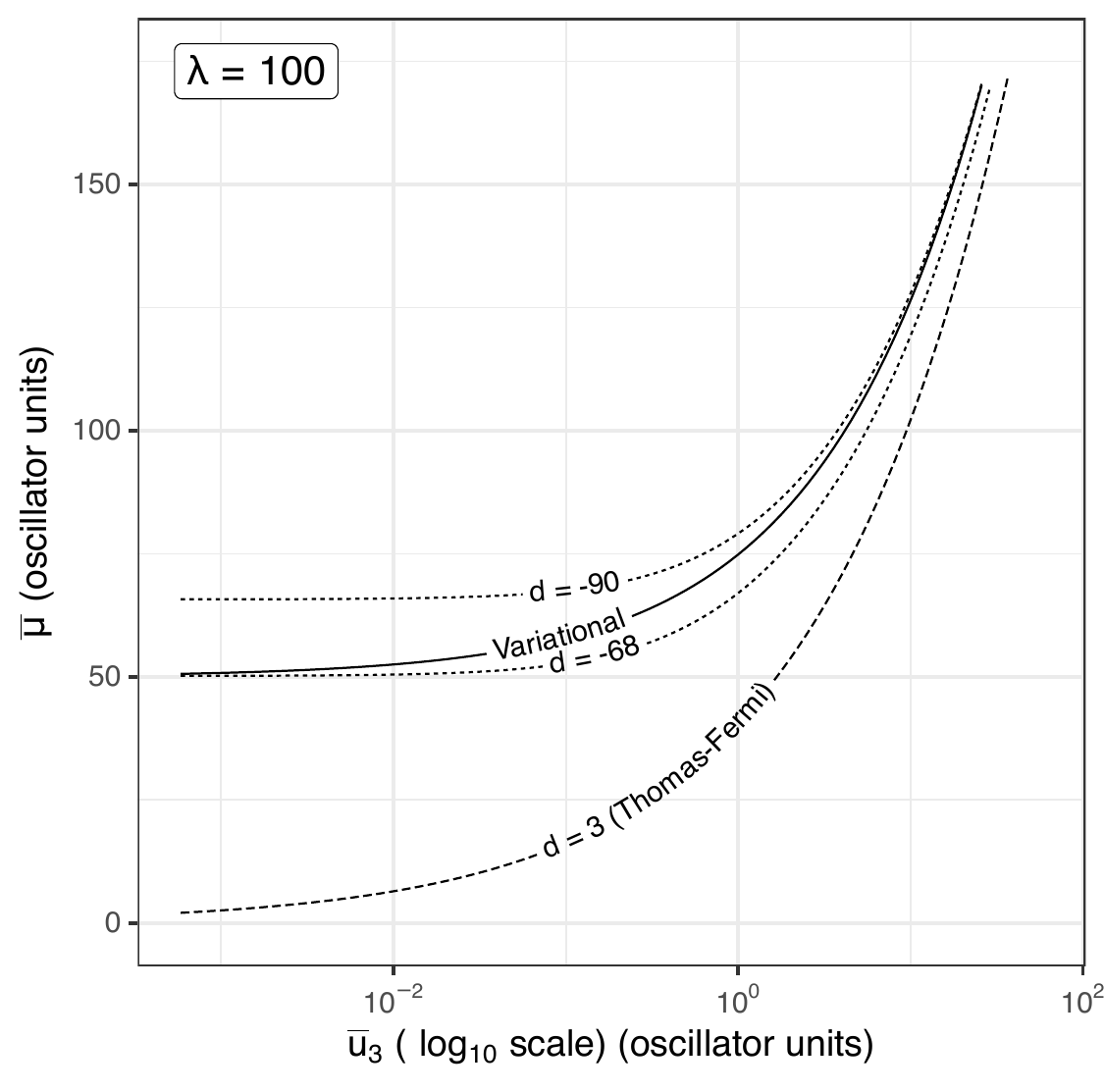}
  \label{fig:diskGeometryMuComparison}
\end{subfigure}
\caption{\textcolor{black}{Zeroth-order approximation of the chemical potential $\bar{\mu}$ (Eq. \ref{eq:3DMuDasUnits} where $\kappa = D + 2|m| - d$) for the $3D$ ground state ($D=3$, $m=0$) compared with the variational approximations from Ref. \cite{Das02} in geometric-mean oscillator units. We test different values for $d$ in the zeroth order approximation, where the value $d=D+2|m|$ is equivalent to the Thomas-Fermi approximation ($d=3$ in this case). For condensate parameters, we use $N=1000$, $m=0$, and a range of interaction strengths $\bar{u}$ in geometric-mean oscillator units. For anistropy, we use $\lambda = 1/100$ (cigar limit, left) and $\lambda=100$ (disk limit, right).}}
\label{fig:muComparison}
\end{figure}

\textcolor{black}{
To more easily compare with Ref. \cite{Das02} and make the effect of $\lambda$ more explicit, we convert our earlier results in transverse oscillator units (based on confinement $\omega_\perp$) to geometric mean oscillator units (based on $\omega=\sqrt[3]{\omega_\perp^2\omega_z}$ in $3D$). Note Ref. \cite{Das02} uses $\gamma$ for anisotropy, which is $1/\lambda$. Generalizing the $\omega$ geometric mean to $D$ dimensions gives}
\begin{equation}\label{eq:generalDomega}
\textcolor{black}{\omega = (\omega_r^{D-1}\omega_z)^{1/D} = \lambda^{1/D} \omega_\perp},
\end{equation}
\textcolor{black}{which we use to convert between generalized geometric mean oscillator units (tilde) and generalized transverse oscillator units (bars). Eq. (\ref{eq:generalDomega}) gives $\bar{\mu}=\lambda^{1/D} \tilde{\mu}$ and $\bar{\alpha} = \lambda^{-1/D}\tilde{\alpha}$. Due to differences in the coupling constant and normalization used in Ref. \cite{Das02} for $3D$, we define the conversion for the coupling constant to be $\bar{u}_D=2\lambda^\frac{2-D}{2D}\Omega_{D-1}\tilde{u}_D$.}

\textcolor{black}{In the geometric mean oscillator units, the general-$D$ zeroth-order equation for the chemical potential (Eq. \ref{eq:generalDMu}) becomes}
\textcolor{black}{\begin{eqnarray}
\label{eq:generalDMuDasUnits}
\frac{2(2\tilde{\mu})^\frac{D+2}{2}}{3\tilde{u}_D} \biggl[\bigg(\frac{\tilde{\alpha}}{\lambda^{1/D}}\bigg)^{D-1} \frac{(D+3)}{2(D-1)(D-3)}
-\bigg(\frac{\tilde{\alpha}}{\lambda^{1/D}}\bigg)^2\frac{3\sqrt{\pi}\Gamma(\frac{D-1}{2})}{4(D-3)\Gamma(\frac{D}{2})} &&\nonumber\\
+ \frac{\sqrt{\pi}\Gamma(\frac{D+1}{2})}{2(D-1)\Gamma(\frac{D+2}{2})} - \frac{\sqrt{\pi}\Gamma(\frac{D+3}{2})}{2(D+1)\Gamma(\frac{D+4}{2})} \biggr] &&~\approx N,
\end{eqnarray}}
\textcolor{black}{and the 3$D$ limit (corresponding to Eq. \ref{eq:3DMu}) becomes}
\textcolor{black}{\begin{eqnarray}\label{eq:3DMuDasUnits} \frac{2(2\tilde{\mu})^\frac{5}{2}}{3\tilde{u}_3} \biggl[ \frac{1}{5}+\frac{3}{2}\biggl(\frac{\tilde{\alpha}}{\lambda^{1/3}}\biggr)^2 \biggl( \ln{\bigg(\frac{\tilde{\alpha}}{2{\lambda^{1/3}}}\bigg)} + \frac{2}{3} \biggr) \biggr] \approx N.
\end{eqnarray}} 
\textcolor{black}{In these units, the effect of $\lambda$ moves from a global prefactor to multiple coefficients that delineate the relative effect of anisotropy on the chemical potential. Similarly, the $2D$ limit is}
\textcolor{black}{
\begin{eqnarray}\label{eq:2DMuDasUnits} 
\frac{2(2\tilde{\mu})^2}{3 \tilde{u}_2}\bigg[ \frac{3\pi}{4}\biggl(\frac{\tilde{\alpha}}{\lambda^{1/2}}\biggr)^2 - \frac{5}{2}\biggl(\frac{\tilde{\alpha}}{\lambda^{1/2}}\biggr) + \frac{3\pi}{16}\bigg]  \approx N.
\end{eqnarray}}
\textcolor{black}{As the BEC is flattened about the $z$-axis ($\lambda \gg 1$, disk), Eq. (\ref{eq:generalDMuDasUnits}) becomes closer to the Thomas-Fermi approximation (Eq. \ref{eq:generalDMuThomasFermi}) as the centrifugal vortex kinetic energy becomes less important. However, physically, the axial kinetic energy becomes more important when $\lambda \gg 1$. As the BEC is stretched along the z-axis ($\lambda \ll 1$, cigar), our transverse correction to the kinetic energy is more significant and physically the transverse kinetic energy is also more important. }


\textcolor{black}{The optimal zeroth-order value of $d$ in $\kappa$ is $\lambda$-dependent (Fig. \ref{fig:muComparison}). The value of $d$ affects the contribution of the centrifugal term to the kinetic energy through the scaled units. For the value $d=3$, the zeroth-order approximation reduces to Thomas-Fermi (zero kinetic energy) when $D=3$ and $|m|=0$. For other (non-Thomas Fermi) values of $d$, the zeroth-order approximation performs better than Thomas Fermi, especially at low interaction strength, because the zeroth order includes zero-point energy from the centrifugal term. For $\lambda=1/100$, the optimal solution lies between $d=1$ and $d=2$, and the zeroth-order density is more appropriate because of the transverse kinetic energy contribution. For $\lambda=100$, larger, more negative values of $d$ are needed to add more transverse kinetic energy to compensate for the neglect of the axial kinetic energy.}

\section{Summary and Conclusions}\label{sec:conclusions}

We used the hypercylindrical Gross-Pitaevskii equation (Eq. \ref{eq:generalDGPE}) and dimensional perturbation theory (DPT) to investigate vortices in $D$-dimensional Bose-Einstein condensates. The perturbation parameter $\delta=1/\kappa$ ($\kappa=D+2|m|-d$) involves the spatial dimension $D$, axial vorticity $|m|$ of the condensate and a parameter $d$ that controls the kinetic energy contribution at zeroth-order. With the zeroth order, we derived semi-analytical results for arbitrary $D$ and we studied the effect of dimensionality on the density and other properties of the condensate. For example, for higher $D$ the density is more sharply peaked and the vortex core pushes outward (Fig. \ref{fig:densityAtDiffD}) due to the large-$D$ effect of the centrifugal term. We analyzed the effect of anisotropy in the cylindrical geometry as a surrogate for effective lower dimension in an isotropic system, and we derived higher-dimensional vortex properties, which have potential applications in the emerging field of synthetic dimensions. 


Introducing an additional expansion in $\textcolor{black}{\bar{\alpha}} = \kappa/(4\bar{\mu})$ (Eq. \ref{eq:alpha}), we found general-$D$ expressions for approximating the chemical potential, energy per particle, and critical vortex velocity (Eqs. \ref{eq:generalDMu}, \ref{eq:generalDEnergy}, \ref{eq:critVelocity}). These expressions extend the results for $3D$ axially symmetric condensates in \cite{Sinha97} to arbitrary $D$. We let $d=3$ in $\kappa$ for exact concordance with $D=3$ in Ref. \cite{Sinha97}. In order for the inner and outer radii of the condensate to be real in the $\delta \to 0$ limit of DPT, we found a $D$-dependent lower bound on the chemical potential ($\bar{\mu}\ge \kappa/2$), which may be like a $D$-dependent zero point energy. We also solved the normalization integral using the zeroth-order approximation without expanding in $\textcolor{black}{\bar{\alpha}}$ to provide a more exact approximation for $\bar{\mu}$  (Eq. \ref{eq:generalDMuNoAlpha}). We demonstrated that our zeroth-order approximation for the chemical potential simplifies to the $D$ dimensional Thomas-Fermi approximation (Eq. \ref{eq:generalDMuThomasFermi}) when we let $d=D+2|m|$ in the $\kappa$ parameter. 

\textcolor{black}{Extending the spherical symmetry results from Ref. \cite{McKinney02} to axial symmetry allowed us to study anisotropy effects such as the transition between $D=3$ and $D=1$ as the condensate is compressed or stretched along the $z$-axis using $\lambda$. 
Converting from transverse to geometric mean oscillator units, similar to Ref. \cite{Das02}, allowed us to see the limiting effects of the anisotropy on the zeroth-order chemical potential (Eq. \ref{eq:3DMuDasUnits}). We also compared the zeroth-order approximation to the ground state variational approach in \cite{Das02} for different regimes of anisotropy (Fig. \ref{fig:muComparison}) and found that missing kinetic energy from the zeroth order can be added by modifying $d$ in the perturbation parameter. This suggests that $d$ could be an effective variational parameter in a trial wave function.   }


\textcolor{black}{The zeroth-order chemical potential increases with anisotropy or interaction strength for a given $D$ (see $\bar{\mu}_D$-curves in Figs. \ref{fig:muVersusu} and \ref{fig:muVersusLambda}). For large anisotropy or interaction, higher dimensions have lower chemical potential, which is driven by the $2/(D+2)$ $D$-dependence in the exponent of the general Thomas-Fermi limit (Eq. \ref{eq:generalDMuThomasFermi}). For weak interaction, the Thomas-Fermi contribution goes to zero and the harmonic oscillator energy dominates, and we observe that the chemical potential increases with increasing D, consistent with a noninteracting harmonic oscillator. This results in the crossings between the $\bar{\mu}_D$-curves for different $D$. As $D$ increases, the density is squeezed and the kinetic energy increases; however, for larger interactions, this increase in kinetic energy is counteracted by the decrease in interaction energy with respect to $D$ in the Thomas-Fermi limit. This effect of the increase in interaction strength or anisotropy may cause a higher-dimensional condensate to behave like a lower-dimensional system. More study is needed to determine whether the dimensional crossings are physical or a mathematical artifact, possibly due to the perturbation limit. The change in behavior from decreasing to increasing chemical potential and the crossing of $\bar{\mu}_D$-curves could possibly be tested in synthetic dimensions by tuning the interaction strength or anisotropy. }  

These findings contribute to our understanding of how the chemical potential and density behave in BECs with varying dimensions, interaction strengths, vorticity and anisotropy, offering insights into the properties of lower and higher-dimensional condensates and their potential applications. \textcolor{black}{In addition to manipulating internal and external states, theoretical work with double-well BECs has shown that synthetic dimensions can be given by the number of atoms in each well \cite{ Mumford2024}. That is, the dimension is determined by the relative number of atoms in a two-dimensional Fock space lattice. Our analytical and numerical results for the $D=4$ GPE with vortices may have applications for synthetic dimensions based on Fock lattices.}    

\textcolor{black}{Future work will investigate potential applications of our hypercylindrical vortex formalism for gravitational models and rotating black holes. The authors in Ref. \cite{Tononi2024} explore how curved geometry affects the structure, stability and dynamics of quantum vortices in atomic BECs using the GPE in cylindrical coordinates. The mathematical techniques developed for curved 2D manifolds could be extended to study vortices in hypercylindrical symmetry with possible analogous gravitational models. Our DPT results involve a $D-1$ hypersphere embedded in $D$ dimensions analogous to a holographic duality mapping of spacetime in $D$ spatial dimensions to a boundary in $D-1$ spatial dimensions. Holographic duality from string theory involves the gravitational physics of anti-de Sitter spacetimes in $D$ spatial dimensions that correspond to a large N conformal field theory on the boundary ($D-1$ spatial dimensions) at strong coupling. The focus of the current study is general-D approximations for the GPE quantum system; however, comparison to the results obtained through holographic means could suggest interesting gravitational physics.} 

Finally, an important future work will be to extend these zeroth-order results to first and higher order in $\delta$ using DPT and to use $d$ as a variational parameter. Higher-order corrections will likely adjust for less accurate choices of $d$ at zeroth order, but a variational approach to optimize $d$ could provide a good $\lambda$-dependent starting point for zeroth order. Higher-order corrections in DPT typically use a harmonic oscillator basis. Similarly, extending the zeroth-order results using a variational approach would benefit from a harmonic oscillator trial wave function that includes kinetic energy in the axial direction combined with the zeroth-order density that includes transverse kinetic energy due to the vortex or due to higher dimensions. 


\newpage
\appendix
 
\appendix\newpage\markboth{Appendix}{Appendix}
\renewcommand{\thesection}{\Alph{section}}
\numberwithin{equation}{section}
 
\section{Full zeroth-order large-$\kappa$ chemical potential and energy approximations using hypergeometric functions}
\label{First_Appendix}

Instead of using an expansion in $\textcolor{black}{\bar{\alpha}}$ to compute the normalization integral (Eq. \ref{eq:generalDNormalization}), one can solve the integral exactly, still using the zeroth-order $\psi$ approximation (Eq. \ref{eq:zerothOrderDensity}). This results in a more exact normalization condition that can be solved numerically for $\bar{\mu}$:

\begin{eqnarray} \label{eq:generalDMuNoAlpha}
\frac{4\Omega_{D-1}}{3\lambda} \frac{(2\bar{\mu})^\frac{D+2}{2}}{\bar{u}_D} \frac{i\sqrt{\pi}|\textcolor{black}{\bar{\alpha}}|}{2^{\frac{D+2}{2}}}\cdot \nonumber\\
\Biggl[\left(\sqrt{1-4\textcolor{black}{\bar{\alpha}}^2}+1\right)^{\frac{D-4}{2}} \Biggl(\frac{4\textcolor{black}{\bar{\alpha}}^2 \Gamma\left(\frac{D-2}{2}\right){}_2F_1\left(-\frac{1}{2},\frac{D-4}{2};\frac{D-1}{2};\frac{\sqrt{1-4 \textcolor{black}{\bar{\alpha}} ^2}+1}{1-\sqrt{1-4 \textcolor{black}{\bar{\alpha}} ^2}}\right)}{(D-4) \Gamma\left(\frac{D-1}{2}\right)} \nonumber\\+ \frac{\left(\sqrt{1-4 \textcolor{black}{\bar{\alpha}} ^2}+1\right)^2 \Gamma\left(\frac{D+2}{2}\right) {}_2F_1\left(-\frac{1}{2},\frac{D}{2};\frac{D+3}{2};\frac{\sqrt{1-4 \textcolor{black}{\bar{\alpha}} ^2}+1}{1-\sqrt{1-4 \textcolor{black}{\bar{\alpha}} ^2}}\right)}{D\,\Gamma\left(\frac{D+3}{2}\right)} \nonumber\\
-\frac{2 \left(\sqrt{1-4 \textcolor{black}{\bar{\alpha}} ^2}+1\right) \Gamma \left(\frac{D}{2}\right) {}_2F_1\left(-\frac{1}{2},\frac{D-2}{2};\frac{D+1}{2};\frac{\sqrt{1-4 \textcolor{black}{\bar{\alpha}} ^2}+1}{1-\sqrt{1-4 \textcolor{black}{\bar{\alpha}} ^2}}\right)}{(D-2) \Gamma \left(\frac{D+1}{2}\right)}\Biggr) &&\nonumber\\
-\left(1-\sqrt{1-4 \textcolor{black}{\bar{\alpha}} ^2}\right)^{\frac{D-4}{2}} \Biggl(\frac{4 \textcolor{black}{\bar{\alpha}} ^2 \Gamma \left(\frac{D-2}{2}\right) {}_2F_1\left(-\frac{1}{2},\frac{D-4}{2};\frac{D-1}{2};\frac{1-\sqrt{1-4 \textcolor{black}{\bar{\alpha}} ^2}}{\sqrt{1-4 \textcolor{black}{\bar{\alpha}} ^2}+1}\right)}{(D-4) \Gamma \left(\frac{D-1}{2}\right)} &&\nonumber\\
+\frac{\left(1-\sqrt{1-4 \textcolor{black}{\bar{\alpha}} ^2}\right)^2 \Gamma \left(\frac{D+2}{2}\right) {}_2F_1\left(-\frac{1}{2},\frac{D}{2};\frac{D+3}{2};\frac{1-\sqrt{1-4 \textcolor{black}{\bar{\alpha}} ^2}}{\sqrt{1-4 \textcolor{black}{\bar{\alpha}} ^2}+1}\right)}{D\,\Gamma \left(\frac{D+3}{2}\right)} &&\nonumber\\
-\frac{2 \left(1-\sqrt{1-4 \textcolor{black}{\bar{\alpha}} ^2}\right) \Gamma \left(\frac{D}{2}\right) {}_2F_1\left(-\frac{1}{2},\frac{D-2}{2};\frac{D+1}{2};\frac{1-\sqrt{1-4 \textcolor{black}{\bar{\alpha}} ^2}}{\sqrt{1-4 \textcolor{black}{\bar{\alpha}} ^2}+1}\right)}{(D-2) \Gamma \left(\frac{D+1}{2}\right)}\Biggr)\Biggr] &&\approx N
\end{eqnarray}
As before, $\Omega_{D-1}$ is the angular integral over a $(D-1)$ dimensional sphere, $\lambda=\frac{\omega_z}{\omega_{\perp}}$ is the anisotropy parameter and $N$ is the number of particles. Both chemical potential $\bar{\mu}$ and interaction strength $\bar{u}_D$ are in transverse oscillator units. The presence of $\textcolor{black}{\bar{\alpha}}=(D + 2|m| - 3)/(4\bar{\mu})$ comes from the definition used in Eq. (\ref{eq:zerothOrderDensityAlpha}), even though here we do not expand Eq. (\ref{eq:generalDMuNoAlpha}) in $\textcolor{black}{\bar{\alpha}}$. 

${}_2F_1$ is the hypergeometric function, which can be regularized using the  $\Gamma$ functions in the denominators to remove singularities. The quantity $i$ is the imaginary unit, but for $0 < \textcolor{black}{\bar{\alpha}} < 1/4$ the expression in Eq. (\ref{eq:generalDMuNoAlpha}) is real. Despite the terms $\frac{1}{D}$, $\frac{1}{D-2}$ and $\frac{1}{D-4}$ in Eq. (\ref{eq:generalDMuNoAlpha}), the limits at $D=0$, $2$ and $4$ do exist. 

\section{Higher-order $\textcolor{black}{\bar{\alpha}}=\kappa/(4\mu)$ expansions of the normalization condition integrand}
\label{Second_Appendix}

The normalization condition in Eq. (\ref{eq:generalDMu}) uses a 2nd-order expansion in $\textcolor{black}{\bar{\alpha}}$, but higher-order expansions are possible. Since the condensate density (Eq. \ref{eq:zerothOrderDensityAlpha}) is a function of $\textcolor{black}{\bar{\alpha}}^2$, the odd terms in the $\textcolor{black}{\bar{\alpha}}$ expansion of the normalization integrand in Eq. (\ref{eq:generalDNormalizationzInt}) are $0$. The 4th-order alpha expansion of the normalization integrand in Eq. (\ref{eq:generalDNormalizationzInt}) results in the normalization condition 
\begin{eqnarray}\label{eq:generalDMu4thOrderAlpha}
\frac{4\Omega_{D-1}(2\bar{\mu})^{\frac{D+2}{2}} }{3 \bar{u}_D \lambda} \biggl[\frac{\textcolor{black}{\bar{\alpha}}^{D-1}\left(D-9)(D+13\right) }{8 (D-5) (D-3) (D-1)}+\frac{3\textcolor{black}{\bar{\alpha}}^4\sqrt{\pi }  \Gamma \left(\frac{D-3}{2}\right)}{8 (D-5) \Gamma \left(\frac{D-4}{2}\right)}&&\nonumber\\
+\frac{3\textcolor{black}{\bar{\alpha}}^2\sqrt{\pi }}{2} \left(\frac{\Gamma \left(\frac{D+1}{2}\right)}{(D-1) \Gamma \left(\frac{D}{2}\right)}-\frac{\Gamma \left(\frac{D-1}{2}\right)}{(D-3) \Gamma \left(\frac{D-2}{2}\right)}\right)&&\nonumber\\
+\sqrt{\pi } \left(\frac{\Gamma \left(\frac{D+1}{2}\right)}{(D-1) \Gamma \left(\frac{D}{2}\right)}-\frac{2 \Gamma \left(\frac{D+3}{2}\right)}{(D+1) \Gamma \left(\frac{D+2}{2}\right)}+\frac{\Gamma \left(\frac{D+5}{2}\right)}{(D+3) \Gamma \left(\frac{D+4}{2}\right)}\right)\biggr]&&~=N,
\end{eqnarray}
where, as in Eq. (\ref{eq:generalDMu}), $\bar{\mu}$ is the chemical potential, $\lambda=\frac{\omega_z}{\omega_{\perp}}$ is the anisotropy parameter, $\bar{u}_D$ is the interaction strength and $N$ is the number of particles. The 6th-order expansion of the normalization integrand results in the normalization condition  
\begin{eqnarray}\label{eq:generalDMu6thOrderAlpha}
\frac{4\Omega_{D-1}(2\bar{\mu})^{\frac{D+2}{2}}}{3 \bar{u}_D \lambda } \biggl[\frac{\textcolor{black}{\bar{\alpha}}^{D-1}\left(D-9) (D^2+12D-109\right) }{16 (D-7) (D-5) (D-3) (D-1)}-\frac{\textcolor{black}{\bar{\alpha}}^5}{80}&&\nonumber\\
-\frac{\textcolor{black}{\bar{\alpha}}^6\sqrt{\pi}\Gamma \left(\frac{D-5}{2}\right)}{8(D-7)\Gamma \left(\frac{D-8}{2}\right)}+\frac{3\textcolor{black}{\bar{\alpha}}^4\sqrt{\pi}}{4} \left(-\frac{\Gamma \left(\frac{D-3}{2}\right)}{(D-5) \Gamma \left(\frac{D-6}{2}\right)}+\frac{\Gamma \left(\frac{D-1}{2}\right)}{(D-3) \Gamma \left(\frac{D-4}{2}\right)}\right)&&\nonumber\\
+3\textcolor{black}{\bar{\alpha}}^2\sqrt{\pi } \left(\frac{\Gamma \left(\frac{D-1}{2}\right)}{(D-3) \Gamma \left(\frac{D-4}{2}\right)}-\frac{2 \Gamma \left(\frac{D+1}{2}\right)}{(D-1) \Gamma \left(\frac{D-2}{2}\right)}+\frac{\Gamma \left(\frac{D+3}{2}\right)}{(D+1) \Gamma \left(\frac{D}{2}\right)}\right)&&\nonumber\\
+2 \sqrt{\pi } \Biggl(-\frac{\Gamma \left(\frac{D+1}{2}\right)}{(D-1) \Gamma \left(\frac{D-2}{2}\right)}+\frac{3 \Gamma \left(\frac{D+3}{2}\right)}{(D+1) \Gamma \left(\frac{D}{2}\right)}&&\nonumber\\
-\frac{3 \Gamma \left(\frac{D+5}{2}\right)}{(D+3) \Gamma \left(\frac{D+2}{2}\right)}+\frac{\Gamma \left(\frac{D+7}{2}\right)}{(D+5) \Gamma \left(\frac{D+4}{2}\right)}\Biggr)\biggr] &&~=N.
\end{eqnarray}
Note the fourth order (Eq. \ref{eq:generalDMu4thOrderAlpha}) approximation has a pole at $D=3$ and so the second order approximation is more appropriate for $D=3$. However, the fourth order approximation has a limit for $D=5$. Similarly, the sixth order (Eq. \ref{eq:generalDMu6thOrderAlpha}) approximation has poles at $D=3,5$ and a limit for $D=7$. For Eq. (\ref{eq:generalDMu4thOrderAlpha}) and Eq. (\ref{eq:generalDMu6thOrderAlpha}), the $D=1$ limit exists because of the $\Gamma((D-1)/2)$ in $\Omega_{D-1}$.

\textcolor{black}{\section*{Acknowledgments}
We would like to thank Dr. Casey Cartwright for his helpful comments on the manuscript. We would like to acknowledge the support of the National Science Foundation, Award No. 1911370.}
\section*{References}
\bibliography{gpe_dpt0.bib}
\end{document}